\documentclass [a4paper]{JHEP3}
\usepackage{epsfig}
\title{%
\protect\vspace{5mm}
Flavour violation with a single generation
}
\author{J.-M.~Fr\`ere$^{1}$, M.V.~Libanov$^{1,2}$, E.Ya.~Nugaev$^{2}$, and
S.V.~Troitsky$^{2}$\\
$^{1}$~Service de Physique Th\'{e}orique, CP 225,\\
  Universit\'{e} Libre de Bruxelles, B--1050, Brussels, Belgium;\\
$^{2}$~Institute for Nuclear Research of the Russian Academy of
Sciences,\\
60th October Anniversary Prospect 7a, Moscow 117312 Russia\\
E-mail: \email{frere@ulb.ac.be, ml@ms2.inr.ac.ru,
emin@ms2.inr.ac.ru, st@ms2.inr.ac.ru} }

\preprint{ULB-TH-03-28}
\abstract{
We calculate probabilities of flavour violating processes mediated by
Kalutza-Klein modes of gauge bosons in a model where three generations of
the Standard Model fermions arise from a single generation in (5+1)
dimensions. We discuss a distinctive feature of the model: while the
processes in which the generation number $G$ changes are strongly
suppressed, the model is constrained by those with $\Delta G=0$, for
instance $K\to \mu ^\pm e^\mp$. The bound on the size of the extra
dimensions is $1/R \gtrsim 64$~TeV.}
\begin{document}
\section{Introduction.}
\label{sec:intro}
In models with more than four space-time dimensions,
some of long-standing problems of particle physics acquire elegant
solutions (see Ref.~\cite{RuUFN} for a review). In particular, in the
frameworks of ``large extra dimensions'' \cite{LED}, the models have been
suggested \cite{Libanov:2000uf,Frere:2000dc} and studied
\cite{Frere:2001ug,Libanov:2002tm} where three generations of the Standard
Model fermions appear as three zero modes localized in the
four-dimensional core of a defect with topological number three. When both
fermions and Higgs boson are localized on the brane, the overlaps of their
wave functions may result in a hierarchical pattern of fermion masses and
mixings \cite{Schmaltz}. This occurs naturally in the models under
discussion \cite{Libanov:2000uf}. To incorporate four-dimensional gauge
fields, a compactified version of the model has been developed
\cite{FLNT}. There, fermions and scalar fields are localized in the core
of a (5+1)-dimensional vortex with winding number three, and two extra
dimensions form a sphere accessible for (non-localized) gauge bosons. The
zero modes of the gauge bosons are independent from coordinates on the
sphere, while higher modes have non-trivial profiles, and hence different
overlaps with fermionic zero modes. Since in our model three
four-dimensional families appear from a single six-dimensional generation,
one can expect flavour violation in the effective four-dimensional theory.
Here, we study the specific pattern of these flavour-violating effects,
which could distinguish the models of this class from other
extra-dimensional models by signatures in rare processes at low energies.

In Sec.~\ref{sec:gauge}, we briefly review the model of Ref.~\cite{FLNT}
and discuss the decomposition of gauge fields on a six-dimensional
manifold. The couplings of the gauge modes to fermions are calculated in
Sec.~\ref{sec:gauge+fermi}. We study specific flavour-violating processes
in Sec.~\ref{sec:processes} and conclude in Sec.~\ref{sec:concl} with a
description of signatures specific for the given class of models. In
Appendices, the notations are summarized together with technical details
and explicit formulae required for calculations.

\section{Gauge bosons on $M^4\times S^2$.}
\label{sec:gauge}
We study the model initially formulated in
Ref.~\cite{Libanov:2000uf} and developed with a simpler field
content in Ref.~\cite{Frere:2000dc}.
The model has been compactified
on $M^4\times S^2$, a
product of our four-dimensional Minkowski space and a two-dimensional
sphere, in Ref.~\cite{FLNT} (see Appendix \ref{sec:def} for
notations). In what follows we will argue that the choice of the manifold
is not important for our principal conclusions. The extra dimensions can
even be infinitely large, as, for instance, in Ref.~\cite{SHAPOSHNIKOV},
where well localized gauge-boson zero modes appear. In this case, the
role of the radius $R$ of the $S^2$ sphere is taken by a
typical size of the localized gauge zero modes but not by a size of extra
dimensions.

The interaction of vector-like counterparts of the fermions of one
Standard Model generation with the vortex field of the Abelian Higgs model
results in $k$ chiral zero modes of each fermion localized in the
four-dimensional core of the vortex; $k$ is the winding number of the
vortex and is equal to three in our case. These three zero modes of a
single six-dimensional fermion represent the corresponding
four-dimensional fermions of three generations; the zero modes are
linearly independent and hence have different windings in $\varphi$.
Detailed descriptions of the model can be found in
Refs.~\cite{Libanov:2000uf,Frere:2000dc,FLNT}; here, we only outline the
setup and introduce some notations.

The vortex is formed by a scalar field, which extends to a typical size
$R\theta_\Phi$ from the origin, and a gauge field of size $R\theta_A$.
Apart from these two fields whose non-trivial profiles have a topological
origin, there is also the Standard Model Higgs doublet $H$ which, due to
the interaction with the vortex scalar field, also develops a non-trivial
profile (see Ref.~\cite{WITTEN}): it is non-zero inside the core of the
vortex, and its vacuum expectation value vanishes in the bulk. The typical
size of $H$ is also $R\theta_\Phi$. Due to the interaction with $H$, the
fermionic zero modes, whose size is $R\theta_A$, aquire small (as compared
to the energy scale of the vortex) masses. The hierarchical fermionic mass
pattern is governed by a small parameter \cite{FLNT}
\[
\delta=\frac{\theta_\Phi}{\theta_A}\sim 0.1.
\]
In what follows we will also assume that $\theta_\Phi<\theta_A\ll 1$.

The gauge bosons of the Standard Model $SU(3)\times SU(2) \times U(1)$
group do not interact directly with the vortex field.\footnote{We neglect
here possible kinetic mixing between the hypercharge and vortex gauge
bosons which is irrelevant for flavour violation.} The interaction of the
$SU(2)\times U(1)$ bosons with the Higgs doublet $H$ ensures the proper
pattern of the electroweak symmetry breaking inside the core of the
 vortex, that is in our usual four-dimensional space.

Let us perform a Kaluza-Klein decomposition of a gauge field. We start
with the  $U(1)$ gauge field ${\cal A}_A$ whose action on $M^4\times S^2$
is
\[
S=-{1\over 4}\int\!d^6X\,\sqrt{-G}G^{AC}G^{BD}{\cal F}_{AB}{\cal
F}_{CD},
\]
where ${\cal F}_{AB}=\partial_A {\cal A}_B-\partial_B {\cal A}_A$. The
separation of variables in the equations of motion is straightforward in
the gauge
\[
\partial _\mu {\cal A} ^\mu =0,
\]
\[
\frac{1}{\sin\theta }\partial _\theta (\sin\theta {\cal A}_\theta)
+\frac{1}{\sin^2\theta }\partial_ \varphi {\cal A}_\varphi =0
\]
and results in the four-dimensional effective Lagrangian for the
Kaluza-Klein modes,
\[
{\cal L}={\cal L}_V+{\cal L}_S,
\]
where
\begin{equation}
{\cal L}_V=\frac{1}{2}A_\nu \partial _\mu ^2A^\nu +\frac{1}{2}\sum
\limits_{l=1}^{\infty }A_{l,\nu }\left(\partial _\mu ^2+\frac{l(l+1)}{R^2}
\right)A_l^\nu + \sum\limits_{l=1}^{\infty  }\sum
\limits_{m=1}^{l}A^*_{(l,m)\nu }\left(\partial _\mu ^2+\frac{l(l+1)}{R^2}
\right)A_{(l,m)}^\nu,
\label{Eq/Pg1/1:vector}
\end{equation}
\begin{equation}
{\cal L}_S=-\frac{1}{2}\sum \limits_{l=1}^{\infty }B_l\left(\partial _\mu
^2+\frac{l(l+1)}{R^2}\right)B_l-\sum \limits_{l=1}^{\infty }\sum
\limits_{m=1}^{l}B_{l,m}^*\left(\partial _\mu
^2+\frac{l(l+1)}{R^2}\right)B_{l,m} .
\label{Eq/Pg1/2:vector}
\end{equation}
The only massless gauge field $A_\mu$, Kaluza-Klein vector fields
$A_{l,\nu}$, $A_{(l,m),\nu}$ and massive scalar fields $B_l$, $B_{l,m}$
are defined in Appendix \ref{app:decomp}. The massless mode $A_\mu$
represents the four-dimensional gauge field which depends neither on
$\theta$ nor on $\varphi$ in our case.

We turn now to the non-abelian gauge bosons. For the unbroken gauge
symmetry case, the quadratic Lagrangian of a non-abelian vector field
reproduces Eqs.~(\ref{Eq/Pg1/1:vector}), (\ref{Eq/Pg1/2:vector}).
Non-observation of Kalutza-Klein modes at colliders implies that the size
$R$ of extra dimensions should be significantly smaller than inverse
$Z$-boson mass. This fact allows one to treat the electroweak symmetry
breaking perturbatively. In this approach, the impact of the background
Higgs field $H(\theta)$, which is localized in the core of the vortex (see
Ref.~\cite{Frere:2000dc}) for details), on the eigensystem of gauge modes,
is considered as a small perturbation compared to curvature of the sphere.
We demonstrate in Appendix \ref{EW} that the lowest modes of $W^\pm$ and
$Z$ bosons aquire proper masses in this way. The corresponding
eigenfunction is no longer constant. However, since the Higgs background,
$H(\theta)$, is independent of $\varphi$, the perturbed mode does not
depend on $\varphi$ as well. As we will see in Appendix
\ref{app:LgaugeFermi}, it means that this mode does not mediate
flavour-changing processes. Contrary to the unbroken case, the $\theta$
dependence of the lowest mode results in different couplings of the
fermions of three generations to $W^\pm$ and $Z$ bosons, given different
localizations of the fermions in extra dimensions. This effect, however,
is suppresed as compared to the family non-universal interactions of
fermions with non-zero modes of gauge bosons.

\section{Coupling of gauge modes to fermions}
\label{sec:gauge+fermi}

The interaction of a six-dimensional fermion $\Psi$ with the
six-dimensional photon ${\cal A}_A$ is determined by the following term in
the Lagrangian,
\begin{equation}
{\cal L}_6=\sqrt{-G}e_6{\cal A}_A\bar{\Psi }\Gamma ^A(X)\Psi\equiv
\sqrt{-G}e_6{\cal A}_AJ^A,
\label{Eq/Pg3/1:vector}
\end{equation}
where coordinate-dependent Dirac matrices $\Gamma^A(X)$ are defined in
Appendix~\ref{sec:def}. In the zero-mode approximation,
\begin{equation}
\Psi (X)=
\sum \limits_{n=1}^{3}a_n(x)\otimes \left(
\begin{array}{c}
0\\
f_2(\theta ,n)\cdot{\rm e}^{i\varphi \frac{7-2n}{2}}\\
f_3(\theta ,n)\cdot{\rm e}^{i\varphi \frac{1-2n}{2}}\\
0
\end{array}
\right),
\label{Eq/Pg3/2:vector}
\end{equation}
where $a_n(x)$, $n=1,2,3$, are three four-dimensional two-component
spinors which represent three generations of fermions with quantum numbers
of $\Psi$ (see Refs.~\cite{Libanov:2000uf,FLNT} for details). The
functions $f_i$ are normalized as
\begin{equation}
2\pi \int\limits_{0}^{\pi }\!d\theta \sqrt{-G}[f_2(n)^2+f_3(n)^2]=1.
\label{Eq/Pg3/2A:vector}
\end{equation}
It is important to note that the angular momentum in transverse dimensions
(that is, the winding number of a wave function) corresponds to the number
of generation: the index $n$ in (\ref{Eq/Pg3/2:vector}) enumerates the
families. As a result, this number is conserved in the first approximation
(this symmetry is broken by the terms responsible for inter-generation
mixing, see Ref.~\cite{Frere:2000dc}). As we will see below, this feature
results in unusually strong suppression of many flavour-violating
processes.

The effective four-dimensional fermion--gauge Lagrangian is calculated in
Appendix \ref{app:LgaugeFermi}. It can be conveniently rewritten as
\begin{equation}
{\cal L}_4=e\cdot{\rm Tr}({\bf A}^\mu {\bf j^*}_\mu ),
\label{Eq/Pg6/1:vector}
\end{equation}
where
\begin{equation}
{\bf A}^\mu =({\bf A}^\mu )^\dag =\sum \limits_{l=0}^{\infty } \left(
\begin{array}{ccc}
E_{11}^{l,0}A_{l,0}^\mu &E_{12}^{l,1}A_{l,1}^\mu & E_{13}^{l,2}A_{l,2}^\mu \\
E_{21}^{l,1}A_{l,1}^{\mu*}&E_{22}^{l,0}A_{l,0}^\mu &
E_{23}^{l,1}A_{l,1}^\mu \\
E_{31}^{l,2}A_{l,2}^{\mu*}&E_{32}^{l,1}A_{l,1}^{\mu*}& E_{33}^{l,0}
A_{l,0}^\mu
\end{array}
\right),
\label{Eq/Pg6/2:vector}
\end{equation}
\[
j_{mn}^\mu(x)=a_m^\dagger \bar\sigma^\mu a_n,
\]
and $e\equiv {e_6\over \sqrt{4\pi}R}$ is the usual four-dimensional
coupling.

The coupling constants $E_{mn}^{l,n-m}$ are defined and estimated in
Appendix \ref{app:LgaugeFermi},
\begin{equation}
E_{mn}^{l,m-n}\sim \left\{
\begin{array}{lcl}
\displaystyle l^{|m-n|+1/2}\theta _A^{|m-n|} & \mbox{at} &
l\theta_A\ll1,\\
\displaystyle \frac{1}{\sqrt{\theta _A}} & \mbox{at} &
l\simeq\displaystyle\frac{1}{\theta _A},\\
\displaystyle {\rm e}^{-lF(\theta _A)}       & \mbox{at} &
l\theta_A\gg1.
\end{array}
\right.
\label{E}
\end{equation}

We see that the fermions have strongest couplings to the heavy modes
with masses
\[
m_l=\frac{\sqrt{l(l+1)}}{R}\sim \frac{1}{\theta _AR}.
\]
The reason for this is obvious: modes with $l\sim 1/\theta$ have largest
overlaps with fermionic wavefunctions of the size $\theta$
($\theta\approx\theta_A$ in our case);
(lower
modes have larger width in $\theta$ while higher modes oscillate several
times at the width of the fermions). We stress that this feature depends
neither on details of localization of fermions and gauge bosons nor on the
shape and size of extra dimensions.

The fermions $a_n$, which enter the current {\bf j}$_\mu$ and consequently
appear in the Lagrangian (\ref{Eq/Pg6/1:vector}), are the states in the
gauge basis, while physically observed mass eigenstates are their linear
combinations. In particular, the mass matrix of the fermions with quantum
numbers of the down-type quarks is given \cite{Frere:2000dc,FLNT} by
\begin{equation}
M_D= \left(
\begin{array}{ccc}
m_{11}&m_{12}&0\\
0&m_{22}&m_{23}\\
0&0&m_{33}
\end{array}
\right) \propto \left(
\begin{array}{ccc}
\delta ^4&\epsilon \delta ^3&0\\
0&\delta ^2&\epsilon \delta \\
0&0&1
\end{array}
\right),
\label{Eq/Pg9/1:vector}
\end{equation}
where
\[
\delta =\frac{\theta _\Phi }{\theta
_A}\sim\sqrt[4]{\frac{m_{11}}{m_{33}}}\sim\sqrt[4]{\frac{m_b}{m_d}}\sim
0.1,
\]
\[
\epsilon\sim 0.1.
\]
To diagonalize the mass matrix one should use biunitary
transformations,
\[
S^\dag _dM_DT_d=M_D^{\rm diag}.
\]

The  fermions in the mass basis are
\[
Q_n=(S^\dag _d)_{nm}q_m\;,\ \ \ D_n=(T^\dag _d)_{nm}d_m,
\]
where we denoted $a_n$ as $q_n$ for the left-handed and as $d_n$ for
the right-handed down-type quarks. If one rewrites the current {\bf
j}$_\mu$ in terms of the mass eigenstates, then the matrix {\bf A}$^\mu$,
Eq.~(\ref{Eq/Pg6/2:vector}), should be replaced by
\[
\tilde{\bf A}^\mu= S^\dag _d{\bf A}^\mu S_d.
\]
Explicit expressions for $S$, $T$ and $\tilde{\bf A}^\mu$ are given in
Appendix \ref{sec:Mgauge+fermi}.

The interaction of fermions with $W^\pm$ and $Z$ bosons is very similar to
the electromagnetic couplings discussed above. There are two differences:
firstly, the current  ${\bf j}_\mu$ in Eq.~(\ref{Eq/Pg6/1:vector}) is
replaced by the Standard-Model charged and neutral weak currents;
secondly, the gauge eigensystem is modified as discussed in Appendix
\ref{EW}. The latter modification does not change the results
significantly: it is negligible for Kalutza-Klein modes and it does not
result in flavour violation for the lowest mode.

To confront the model with the experimental results, one needs to
calculate the effective four-fermion coupling $g_{mn}$, that is, in each
particular case, to sum up the contributions
\[
g_{mn}^l=e^2\frac{(E_{mn}^{l,m-n})^2}{m_l^2}
\]
for all $l$. A very naive estimate gives, using Eq.~(\ref{E}),
\begin{equation}
g_{mn}\sim e^2l_{\mbox{max}}\cdot R^2\theta _A=\frac{e^2}{\theta _A}\cdot
R^2\theta _A=e^2R^2
\label{Eq/Pg8/4:vector}
\end{equation}
so that
\[
\frac{g_{mn}}{G_F}\sim(M_WR)^2.
\]
This result is supported by more explicit calculations given in Appendix
\ref{sec:4fermi}.

\section{Flavour violating processes}
\label{sec:processes}
We turn now to the study of specific flavour
violating processes which are known to give the strongest constraints on
masses and couplings of new vector bosons. The most stringent bounds
arise
\cite{Cahn:1980kv} from $K^0_L$ -- $K^0_S$ mass difference, forbidden $K$
decays $K^0_L \to \mu e$, $K^+ \to \pi^+ e^- \mu^+$ (see also
Ref.~\cite{Ritchie:1993ua}) and from lepton flavour violating processes
$\mu \to e\gamma$, $\mu \to 3e$ and $\mu\to e$ conversion on a nuclei. We
discuss all these constraints below.

First of all, one should note that without account of inter-generation
mixings, the generation number $G$ is exactly conserved. Indeed, the
integration over $\phi$ in the effective Lagrangian results in the
corresponding selection rules: in Eq.~(\ref{Eq/Pg6/2:vector}), no vector
boson has both diagonal and off-diagonal couplings simultaneously. This
forbids all processes with nonzero change of $G$; the probabilities of the
latters in the full theory are thus suppressed by powers of the
mass-matrix mixing parameter, $(\epsilon \alpha)^{\Delta G}$ ($\alpha$ is
determined in Appendix \ref{sec:Mgauge+fermi}). However, the amplitudes of
processes with $\Delta G=0$ but lepton and quark flavours violated
separately are suppressed only by the mass squared of the Kalutza-Klein
modes. The best studied among these processes are kaon decays $K^0_L \to
\mu e$ and $K^+ \to \pi^+ e^- \mu^+$, forbidden in the Standard Model with
massless neutrinos because of separate conservation of $e$ and $\mu$
lepton numbers\footnote{Amplitude of the $K^0_L \to \mu e$ process due to
non-zero neutrino masses is thirty orders of magnitude smaller than the
best experimental limit \cite{Lang-neutrino}.}. In the rest of this
section, we estimate, in the frameworks of the full theory with mixing,
the size of flavour-violating effects for different values of $\Delta G$.

\subsection{$\Delta G=0$: forbidden kaon decays.}
\label{dg=0}
The best experimental restriction on flavour-violating
processes with $\Delta G=0$ is the branching ratio of $K^0_L \to \mu e$
decay \cite{PDG},
\begin{equation}
\mathop{\rm Br} (K^0_L\to\bar\mu^+ e^-)<B=2.4 \cdot 10^{-12}.
\label{k*}
\end{equation}
The $K^0$ meson is a pseudoscalar, and the decay cannot be mediated by
purely vector interaction of the Kalutza-Klein modes of the photon.
However, the higher modes of the $Z$ boson interact with a $V-A$ current
and contribute to the decay width. From Eq.~(\ref{Eq/Pg11/1:vector}) one
obtains, in particular, the dominant, unsuppressed by $(\epsilon\delta)$,
axial coupling in the four-dimensional Lagrangian,
\[
{g\over 2 \cos\theta_W} \sum\limits_{l=1}^\infty E^{l,1}_{1\,2}
Z^\mu_{l,1} \left(-{1\over 2}\bar s \gamma_\mu \gamma_5 d- {1\over 2}\bar
e \gamma_\mu \gamma_5 \mu - ({1\over 2}-2\sin^2\theta_W)\bar
e \gamma_\mu \gamma_5 \mu \right).
\]
The diagrams for this and other processes are similar to those
given in \cite{Cahn:1980kv,Ritchie:1993ua};
one has to sum over all intermediate Kalutza-Klein modes to obtain the
effective four-fermion coupling in a way similar to
Sec.~\ref{sec:gauge+fermi} or Appendix~\ref{sec:4fermi},
\[
\sum\limits_{l=1}^\infty {(E^{l,1}_{1\,2} R)^2 \over l(l+1)} =\zeta R^2,
\]
where $\zeta\approx 0.4$ is a coefficient which results from numerical
evaluation of the sum.

The partial width of the $K^0_L\to\mu e$ decay is easy to estimate by
comparison to the branching ratio of $K^+\to\mu^+\nu$: in the $m_e\ll
m_\mu$ approximation, the phase volume and $f_K$ factors cancel in the
width ratio,
\[
\mathop{\rm Br}(K^0_L\to\mu e)= {\Gamma(K^0_L\to\mu e) \over
\Gamma(K^0_L\to {\rm all})} = \mathop{\rm Br}(K^+\to\mu^+\nu) {\tau(K^0_L)
\over \tau(K^+)} \left| {\langle \bar\mu e | \bar s  d \rangle \over
\langle \bar\mu \nu | \bar s u  \rangle } \right|^2,
\]
where $\tau(K^0_L)$ and $\tau(K^+)$ are the lifetimes of the corresponding
particles. The interaction responsible for the $K^+\to\mu^+\nu$ decay in
the Standard Model is
\begin{equation}
{g\over 2\sqrt{2}} W^\mu \left( \bar s \gamma_\mu (1+\gamma_5) u \sin
\theta_c + \bar \mu \gamma_\mu (1+\gamma_5) \nu_\mu \right),
\label{k1}
\end{equation}
where $\theta_c$ is the Cabibbo angle, and results
in the four-fermionic matrix element,
\begin{equation}
\langle u \bar s | \bar\mu \nu \rangle_t = {2 g^2 \sin\theta_c \over 8
M^2_W}.
\label{k****}
\end{equation}
Hereafter, we denote as $\langle\dots\rangle_t$ a matrix element
with truncated wave functions and spinorial structure.
Using  Ref.~\cite{Langacker:2000ju}
we obtain theoretical prediction on the branching ratio
of the process $K^0_L\to\mu^+ e^-$, from which
one obtains the following bound on
the size of the sphere,
\begin{equation}
{1\over R} > {M_W \over  \cos\theta_W} \left( {\zeta \over \sin\theta_C}
\right)^{1/2} \left( {\mathop{\rm Br}(K^+\to\mu^+\nu) \over B} \,
{\tau(K^0_L) \over \tau(K^+)} \frac{2\sin^4\theta_W-\sin^2\theta_W+1/4}{2}\right)^{1/4}
\label{k2}
\end{equation}
We use all necessary numerical values from Ref.~\cite{PDG} and obtain the
restriction,
\[
{1\over R}>101\sqrt\zeta ~{\rm TeV}\approx 64~{\rm TeV}.
\]

In a similar way, the width of the decay $K^+\to \pi^+\mu^+ e^-$ can be
compared to one of $K^+\to \pi^0\mu^+\nu$. The relevant interactions are
\[
e\sum\limits_{l=1}^\infty E^{l,1}_{1\,2} A^\mu_{l,1} \left\{ -{1\over
3}\bar s \gamma_\mu d +\bar e \gamma_\mu \mu \right\}
\]
\[
+{g\over 2 \cos\theta_W} \sum\limits_{l=1}^\infty E^{l,1}_{1\,2}
Z^\mu_{l,1} \left\{ \left({2\over 3}\sin^2\theta_W-{1\over 2}\right) \bar
s \gamma_\mu d- \bar e \left(2\sin^2\theta_W -{1\over 2}+{1\over 2}\gamma
_5\right) \gamma_\mu \mu \right\}.
\]
For the decay $K^+\to \pi^0\mu^+\nu$, the relevant interaction is given
by Eq.~(\ref{k1}), and the matrix element $\langle \bar u \bar\mu \nu
|\bar s \rangle_t$ coincides with
Eq.~(\ref{k****}). Together with the limit \cite{PDG}
\[
\mathop{\rm Br}(K^+\to \pi^+\mu^+ e^-)<B_1=2.8 \cdot 10^{-11},
\]
this determines that
\[
{1 \over R}>{M_W\over\cos\theta_W} \left({\frac{\zeta}{2\sin\theta_C}}\right)^{1/2} \left(
{\xi\mathop{\rm Br}(K^+\to \pi^0\mu^+\nu) \over B_1} \right)^{1/4},
\]
where
\[
\xi=(4\sin^2\theta_W/3-1)^2(1+(4\sin^2\theta_W-1)^2)
+\frac{\sin^4\theta_W}{9}(16\cos^2\theta_W)^2.
\]
So, constraint from this decay is
\[
{1\over R}>25~{\rm TeV},
\]
which is less restrictive than Eq.~(\ref{k2}).

\subsection{$\Delta G=1$: lepton flavour violation.}
\label{dg=1}
As we have already noted, the processes with $\Delta G\ne 0$
are suppressed by powers of the mixing parameter $\epsilon \alpha$.
Indeed, it follows from Eq.~(\ref{Eq/Pg11/1:vector}) that these processes
could be mediated by ``diagonal'' vector bosons $A^\mu_{l,0}$, and each
corresponding diagram contains one vertex suppressed by $\epsilon \alpha$.

The analysis of the $\mu\to ee\bar e$ process is very similar to one of
kaon decays. The interaction terms are
\[
e \sum\limits_{l=1}^\infty E^{l,0}_{1\,1} A^\mu_{l,0} \bar
e\gamma_\mu\left[ (\epsilon_L \alpha_L)\mu+e \right]+
\]
\[
{g\over 2 \cos\theta_W} \sum\limits_{l=1}^\infty E^{l,0}_{1\,1}
Z^\mu_{l,0} \left[ \bar e \gamma_\mu \left(\left[2\sin^2\theta_W-{1\over
2}\right]-{1\over 2}\gamma_5\right) \left((\epsilon_L \alpha_L)\mu+e
\right) \right],
\]
we denoted the parameters $\epsilon$ and $\alpha$ of the leptonic
mixing matrix as $\epsilon_L$ and $\alpha_L$. There is also a contribution
to this process (as well as to the $\mu e$-conversion discussed below)
mediated by ``off-diagonal'' bosons $A_{l,1}^\mu$. The contribution has
the same order (suppresed by $\epsilon_L\alpha_L$) and the opposite sign.
We do not suspect, however, that some cancellation of the
diagrams is present.

Following Ref.~\cite{Langacker:2000ju} again,
we obtained width of decay
$$
\Gamma (\mu \to ee\bar e)={G_F^2 m_\mu ^5\over 192 \pi ^3}(m_WR)^4 (\epsilon _l\alpha_S)^2
(\zeta)^2 \frac{1+20 \sin^4\theta_W}{2\cos^4\theta_W},
$$
so the limit is
$$
1/R>96 \sqrt{\epsilon _l\alpha_L\zeta}~{\rm TeV}.
$$
In the leptonic sector, the mixing parameter $\epsilon_L$ is unknown;
however even $\alpha_L\sim\delta_L\sim (m_e/m_\tau)^{1/4}$ results in
additional suppression.

Traditionally, one of the strongest constraints on the masses and
couplings of new vector bosons arises from $\mu e$-conversion on nuclei.
Let us show that, in our model, this bound is not so restrictive.
Adopting the calculation of
Ref.~\cite{Bernabeu},\cite{Kuno:1999jp} to our case, we estimate the relative muon conversion
rate on a nucleus with charge $Z$ and neutron number $N$ as
\[
{\cal R}\equiv {\Gamma_{\rm conv} \over \Gamma_{\rm capt}} =
2{\alpha^3_{\rm QED} m_\mu^5 \over \pi^2 \Gamma_{\rm capt}} {Z_{\rm eff}^4
\over Z} |F(q)|^2
(\epsilon_L\alpha_L)^2 \zeta^2 R^4 \kappa,
\]
where
\[
\kappa= \left(
|(2Z+N)\xi _{Lu}+(Z+2N)\xi _{Ld}|^2 + |(2Z+N)\xi _{Ru}+(Z+2N)\xi _{Rd}|^2
\right),
\]
and
$$
\xi _{Ld}=2 \sin^2\theta_W/3+(-1/2+\sin^2\theta_W)(-1/2+2\sin^2\theta_W/3)/\cos^2\theta_W\approx 0.275;
$$
$$
\xi _{Rd}=2 \sin^2\theta_W/3+\sin^2\theta_W(-1/2+2\sin^2\theta_W/3)/\cos^2\theta_W\approx 0.050;
$$
$$
\xi _{Lu}=-4 \sin^2\theta_W/3+(-1/2+\sin^2\theta_W)(1/2-4\sin^2\theta_W/3)/\cos^2\theta_W\approx -0.374;
$$
$$
\xi _{Ru}=-4 \sin^2\theta_W/3+\sin^2\theta_W(1/2-4\sin^2\theta_W/3)/\cos^2\theta_W\approx -0.249.
$$
For the titanium nuclei \cite{Bernabeu}, the muon capture rate
$\Gamma_{\rm capt}\approx 2.6 \cdot 10^6~{\rm s}^{-1}$, the effective
charge $Z_{\rm eff}\approx 17.6$, the nuclear form factor $|F(q)|\approx
0.54$, $Z=22$, $N=26$ and the strongest limit \cite{PDG} is
\[
{\cal R}<{\cal R}_1\approx 4.3 \cdot 10^{-12}.
\]
We obtain the constraint on $R$ and $(\epsilon_L A_L)$ from the
non-observation of $\mu e$-conversion on nuclei,
\[
{1\over R}> 124 \sqrt{\epsilon_L \alpha_L} {\rm TeV}\approx 12~{\rm TeV}
\]

The bound from $\mu\to e\gamma$ decay is further suppressed by a loop
factor.

\subsection{$\Delta G=2$: $K_L-K_S$ mass difference and CP violation in kaons.}
\label{dg=2}
Non-universal couplings of the gauge bosons would contribute
also to the kaon mass difference $(m_{K_L}-m_{K_S})$ which was measured with a
good accuracy. They also would induce additional CP-violation effects.
In our case, however, these contributions are suppressed.

Indeed, the relevant interaction reads\footnote{ The fermionic wave
functions in this equation do not include the complex phase factors $F_d$,
$F_d^\dag$. These factors redefine (reduce) the phase of $\epsilon _d$
(see Appendix~\ref{sec:Mgauge+fermi} for details).}
\[
-(\epsilon_d \alpha_d) {e\over 3} \sum\limits_{l=1}^\infty
\left(E_{11}^{l,0}-E_{22}^{l,0} \right) A_{l,0}^\mu \bar s\gamma_\mu d
\]
\[
+(\epsilon_d \alpha_d) {g\over 2 \cos\theta_w} \sum\limits_{l=1}^\infty
\left(E_{11}^{l,0}-E_{22}^{l,0} \right) Z_{l,0}^\mu \bar s\gamma_\mu
\left[ \left({2\over 3}\sin^2\theta_w-\frac{1}{2} \right)-{1\over
2}\gamma_5 \right] d
\]
\[
+(\epsilon_d \alpha_d) g_s \sum\limits_{l=1}^\infty
\left(E_{11}^{l,0}-E_{22}^{l,0} \right) G_{l,0}^{\mu i} \bar s\gamma_\mu
{\lambda_i\over 2} d + {\rm h.c.},
\]
where the last term represents the interaction with the Kalutza-Klein
modes of the gluon. The latter contribution dominates over the former two
because of larger coupling $g_s$.

One estimates
the contribution to $\Delta m_K$ from the exchange of the higher modes of
the gluon field as
\begin{eqnarray}
&\Delta 'm_K\approx 2\mbox{Re}\langle K_0|H_{\Delta
G=2}|\bar K_0 \rangle&\nonumber\\
&\!\!\!\!\!\!\!\!=m_Kf_K^2
\displaystyle\frac{8g^2_{S}}{9}\left\{\!(\epsilon_d \alpha_d)^2
+(\epsilon_u \alpha_u)^2
+\left({m_K\over m_s+m_d}\right)^2\!\!\!\!
\epsilon_d \alpha_d\epsilon_u \alpha_u\!\right\}
\!\!\sum\limits_{l=1}^\infty
\left(E_{11}^{l,0}-E_{22}^{l,0}\right)^2 \!\!{R^2\over l(l+1)},&
\label{delta}
\end{eqnarray}
where the matrix element was estimated in the vacuum insertion
approximation (see Ref.~\cite{Masiero} and references therein). We note
that, besides the expected $(\epsilon_d \alpha_d)^{\Delta G}$,
additional suppression factors arise in the four-fermionic interaction
due to the following two reasons. Firstly,
$E_{11}^{l,0}\approx E_{22}^{l,0}$: as it is shown in
Appendix~\ref{app:LgaugeFermi}, $E_{mn}^{l,m-n}$ depends, in the first
approximation, on $|m-n|$ and not on $m$ and $n$ separately.
In the second approximation, in a way similar to
Sec.~\ref{sec:gauge+fermi} or Appendix~\ref{sec:4fermi}, one obtains
\[
\sum\limits_{l=1}^\infty \left(E_{11}^{l,0}-E_{22}^{l,0}\right)^2
{1\over l(l+1)} \sim \theta^2_A\sim 0.01.
\]

The second reason is that the matrices $T_d$ and $S_u$ are almost
diagonal: $\alpha_u\sim\delta^3$ whereas $\alpha_d\sim\delta$
(see Appendix~\ref{sec:Mgauge+fermi}).
This suppresses the third term in the curled brackets in
Eq. (\ref{delta}) down to the order of the first one.
Therefore, for $g_s\approx 1.1$ and all other parameters from
Ref.~\cite{PDG}, we obtain the following limit on $R$,
\[
\begin{array}{c}
\displaystyle
{1\over R}>(\epsilon_d \alpha_d) g_s f_K \theta_A \sqrt{\zeta {8\over 9}
\left(1+\left({m_K \over m_d+m_s}\right)^2\frac{\epsilon_u\alpha_u}
{\epsilon_d\alpha_d}\right)
\frac{m_K}{\Delta m_K}}\\ \\
 =(\epsilon_d
\alpha_d\theta_A)\displaystyle\sqrt{\rule{0pt}{15pt}1+
30\displaystyle\frac{\epsilon_u\alpha_u}
{\epsilon_d\alpha_d}}\cdot1300~{\rm Tev} \approx 1.5~{\rm Tev}.
\end{array}
\]
This value is small compared to the restriction from the limit on
the branching ratio of $\mbox{$K_L\to\mu e$}$ decay. However, it has been
pointed out  in Ref.~\cite{Delgado:1999sv} that
a stronger bound on the size of extra dimensions
may arise due to additional contributions to the
CP-violating parameter $\varepsilon_K$.
In the five-dimensional model of Ref.~\cite{Delgado:1999sv},
this restriction is about an order
of magnitude stronger than one from $\Delta m_K$.
This is not the case in a class of models considered here. Indeed,
\[
\varepsilon_K= \frac{\mbox{Im} \langle K_0|H_{\Delta G=2} |\bar
K_0\rangle} {2\sqrt 2 \Delta m_K}.
\]
The imaginary part arises from the phase of coupling constant of fermions
with higher gauge modes (which
does not necessarily coincide with the phase of CKM matrix).
This phase is also suppressed (see Appendix~\ref{sec:Mgauge+fermi}) due to
the same ``almost'' diagonal structure of the matrix $S_u$. Taking into
account this notion, we obtain the limit
\[
\begin{array}{c}
\displaystyle {1\over R}>(\epsilon_d \alpha_d) g_s f_K \theta_A
\sqrt{\zeta {8\over 9} \left(1+\left({m_K \over m_d+m_s}\right)^2
\frac{\epsilon_u\alpha_u} {\epsilon_d\alpha_d} \right) \frac{m_K}{\Delta
m_K}} \sqrt{\rule{0pt}{20pt} \displaystyle
\frac{\epsilon_u\alpha_u}{\epsilon_d\alpha_d\sqrt 2 \varepsilon_K}}\approx
2.6~{\rm Tev}
\end{array}
\]
Here, the amplification by  $1/\sqrt{\varepsilon_K}$ is present,
similarly to Ref.~\cite{Delgado:1999sv}, but an
additional suppression by a factor of
$\sqrt\frac{\epsilon_u\alpha_u}{\epsilon_d\alpha_d}$
diminishes this effect.

\section{Conclusions}
\label{sec:concl}
The model with a single generation of vector-like
fermions in six dimensions allows one to explain fermionic mass hierarchy
without introducing a flavour quantum number: three families of
four-dimensional fermions appear as three sets of zero modes developed on
a brane by a single multi-dimensional family while the fermionic wave
functions inevitably produce a hierarchical mass matrix due to different
overlaps with the Higgs field profile. The six-dimensional Lagrangian with
one generation contains much less parameters than the effective one. All
masses and mixings of the Standard-Model fermions are governed by a few
parameters of order one. This fact allows for specific phenomenological
predictions. In particular, in a compactified version of the model with
non-localized gauge fields, the Kalutza-Klein modes of the vector bosons
mediate flavour-violating processes studied in this paper. The pattern of
flavour violation is distinctive: contrary to other extra-dimensional
models, processes with change of the generation number $G$ by one or two
units are strongly suppressed compared to other rare processes. For
example, ${\rm Br}(\mu\to \bar{e}ee)/{\rm Br}(K\to\mu^\pm e^\mp)\sim
1/10$. The strongest constraint on the model arises from non-observation
of the decay $K\to\mu^\pm e^\mp$; it requires that the size of the
extra-dimensional sphere (size of the gauge-boson localization) $R$
satisfies $1/R \gtrsim 64$~TeV. The Kalutza-Klein modes of vector bosons
have larger masses, but for large enough $R$, could be
detected indirectly by precision measurements at future linear colliders.
A clear signature of the model would be an observation of $K\to\mu^\pm
e^\mp$ decay without observation of $\mu\to \bar{e}ee$, $\mu\to e\gamma$ and
$\mu e$-conversion at the same precision level.

\acknowledgments
We are indebted to S.~Dubovsky, D.~Gorbunov and V.~Rubakov for numerous
helpful discussions.  This work is supported in part by the IISN
(Belgium), the ``Communaut\'e Fran\c{c}aise de Belgique''(ARC), and the
Belgian Federal Government (IUAP); by RFFI grant 02-02-17398 (M.L., E.N.\
and S.T.); by the Grants of the President of the Russian Federation
NS-2184.2003.2 (M.L., E.N.\ and S.T.), MK-3507.2004.2 (M.L.) and
MK-1084.2003.02 (S.T.);
by INTAS grant YSF 2001/2-129 (S.T.): by fellowships of the
``Dynasty'' foundation (awarded by the Scientific Council of ICFPM)
(E.N. and S.T.) and the Russian Science Support Foundation (M.L. and S.T.).

\appendix \section{Notations.}
\label{sec:def}
The metric $G_{AB}$ of $M^4\times S^2$  is determined by
\[
ds^2=G_{AB} dX^AdX^B=\eta_{\mu \nu }dx^\mu dx^\nu -R^2(d\theta
^2+\sin^2\theta d\varphi ^2),
\]
where $\eta_{\mu \nu }={\rm diag}(+,-,\ldots ,-)$ is the four dimensional
Minkowski metric, capital Latin indices enumerate the coordinates $X^A$ on
$M^4\times S^2$, $A,B=0,\ldots 5$; Greek indices refer to the coordinates
$x_\mu$ on $M^4$, $\mu ,\nu =0,\ldots ,3$. We reserve lower case Latin
indices $a,b=0,\ldots 5$ for a flat six-dimensional tangent space. The
minimal representation for six-dimensional spinors is eight-component. In
the curved space, the Dirac matrices depend on the coordinates,
\begin{equation}
\Gamma ^A(X)=h_a^A\Gamma ^a,
\label{Eq/Pg3/3:vector}
\end{equation}
where the sechsbein in our case is given by
\begin{equation}
h^a_A=(\delta _\mu ^a,\delta _4^aR,\delta _5^aR\sin\theta )
\label{Eq/Pg3/4:vector}
\end{equation}
and flat space $8\times 8$ Dirac matrices are
\begin{equation}
\Gamma ^\mu =\left(
\begin{array}{cc}
\mbox{\Huge $_0$}&
\begin{array}{cc}
\bar{\sigma }^\mu &0\\
0&\sigma ^\mu
\end{array}\\
\begin{array}{cc}
{\sigma }^\mu &0\\
0&\bar{\sigma} ^\mu
\end{array}
& \mbox{\Huge $_0$}
\end{array}
\right) , \ \ \Gamma ^{4,5}= \left(
\begin{array}{cc}
\mbox{\Huge $_0$}&
\begin{array}{cc}
0 &\pm1\\
1&0
\end{array}\\
\begin{array}{cc}
0&\mp1\\
-1&0
\end{array}
& \mbox{\Huge $_0$}
\end{array}
\right),
\label{Eq/Pg3/5:vector}
\end{equation}
where $\sigma_\mu=(1,\sigma_i)$, $\bar\sigma_\mu=(1,-\sigma_i)$, and
$\sigma_i$ ($i=1,2,3$) are the Pauli matrices.

\section{Decomposition of the gauge field.}
\label{app:decomp}
After separation of variables, the six-dimensional
$U(1)$ gauge field ${\cal A}_A$ is decomposed as
\[
\begin{array}{c}
\displaystyle {\cal A}_\mu(X)= \frac{1}{R}\left(\sum \limits_{l=0}^{\infty
}\sum \limits_{m=-l}^{l}A_{(l,m)\mu }(x)Y_{lm}(\theta ,\phi )\right) \\
\displaystyle \equiv \frac{1}{R}\left(\frac{A_\mu(x)}{\sqrt{4\pi }}+
\sum \limits_{l=1}^{\infty } A_{l,\mu }(x)Y_{l0}(\theta ,\phi ) + \sum
\limits_{l=1}^{\infty }\sum \limits_{m=-l,\, m\neq0}^{l}A_{(l,m)\mu}(x)
Y_{lm}(\theta ,\phi )\right),
\end{array}
\]
\[
{\cal A}_\theta(X) =\frac{1}{\sin\theta }\sum \limits_{l=1}^{\infty }\sum
\limits_{m=-l,\, m\neq
0}^{l}B_{l,m}(x)\frac{|m|}{\sqrt{l(l+1)}}Y_{lm}(\theta ,\phi ),
\]
\[
\begin{array}{c}
\displaystyle {\cal A}_\varphi(X) =i\sin\theta \sum \limits_{l=1}^{\infty
}\sum
\limits_{m\neq0}^{}\frac{|m|}{m}\frac{1}{\sqrt{l(l+1)}}B_{l,m}(x)\partial
_\theta Y_{lm}(\theta,\phi) \\
\displaystyle +\sin\theta \sum \limits_{l=1}^{\infty }B_l(x)
\sqrt{\frac{2l+1}{4\pi}}\sqrt{\frac{(l+1)!}{(l-1)!}}P_l^1(\cos \theta ) ,
\end{array}
\]
where $Y_{lm}(\theta ,\phi )$ are properly normalized,
\[
\int\limits_{0}^{2\pi }\!\int\limits_{0}^{\pi }\!
Y_{l'm'}^*Y_{lm}\sin\theta\, d\varphi\, d\theta =\delta _{l,l'}\delta
_{m,m'},
\]
spherical harmonics,
\[
Y_{lm}=(-1)^{\frac{m+|m|}{2}}\sqrt{\frac{2l+1}{4\pi
}}\sqrt{\frac{(l-|m|)!}{(l+|m|)!}}P_l^{|m|}(\cos \theta ){\rm
e}^{im\varphi },
\]
and $P_l^m(x)$ are adjoint Legendre functions.

\section{Electroweak symmetry breaking.}
\label{EW}
In the six-dimensional theory, only one scalar field, $H$, is
charged under the electroweak group \cite{Frere:2000dc}.  The soliton-like
solution for the Higgs-vortex system breaks the electroweak symmetry, and
non-zero values of four-dimensional masses of $W$ and $Z$ bosons arise.
The classical Higgs profile is independent of $\varphi$, so the lowest
modes of massive gauge bosons in the background of the soliton do not
depend on $\varphi$ as well. As a result, the $Z$ boson itself does not
mediate flavour changing processes.  The masses of the lowest modes can be
calculated by means of the perturbation theory in a small parameter $g^2$.
In the zeroth approximation, the eigenfunctions are constant zero modes of
the Laplace operator, equal to $\frac{1}{\sqrt{4\pi}R}$.  The first-order
correction to the Lagrangian is
\[
\Delta {\cal L}_6 = \sqrt{-G}\frac{1}{2}H^2(\theta)(g_6^2(W^{+\,
A}W^-_A) + (g_6^2+{g'}_6^2) Z^{0\, A} Z^0_A),
\]
where $H(\theta)$ is the configuration for the Higgs field \cite{FLNT},
which can be approximated by a step of width $\theta_\phi$. If we denote
$\int R^2d\theta\sin{\theta}d\varphi H^2(\theta)=v^2/2$, the usual masses
for gauge bosons arise as square root of corrections for eigenvalues,
\[
m_w=gv/2,\ \  m_z=\sqrt {g^2+g'^2}v/2,
\]
where $g$, $g'=g_6/\sqrt{4\pi}R$ and $v$ are electroweak constants of
the Standard Model.  So, the result for masses of the gauge bosons is
reproduced if
\[
H(0)\sim\frac{v}{\sqrt{2\pi}R\theta_\Phi}\sim 10^3\mbox{ TeV}^2
\]
at $R\sim 100$ TeV and $\theta_\Phi\sim 0.01$.

The first correction to the profile of the $Z$-boson mode is
\begin{equation}
(g^2+g'^2)R^2\sum_{l=1}^{l=\infty}\frac{V_{0l}}{l(l+1)}\frac{1}{R}
Y_{l0}(\theta),
\label{appsum}
\end{equation}
where
\[
V_{0l} = \int R^2d\theta\sin{\theta}d\varphi H^2(\theta)
\frac{1}{\sqrt{4\pi}} Y_{l0}\sim v^2\left[
\begin{array}{lcl}
\sqrt{l}&\mbox{at}&l\theta_\Phi\leq1,\\
{\rm e}^{-lF(\theta\Phi)}&\mbox{at}&l\theta_\Phi\gg1.
\end{array}
\right.
\]
The last approximation can be obtained in the same way as in
Appendix~\ref{app:LgaugeFermi}.
The value of $V_{0l}$ is maximal at $l\sim
1/\theta_\Phi$, $V_{0l}/l(l+1) \sim 1/l^{3/2}$, and, therefore, the sum
(\ref{appsum}) is saturated by the lightest modes. Thus, the $Z$-boson
mode recives small, of order of
\[
\frac{1}{R}(g^2+g'^2)(Rv)^2\sim 10^{-4}g^2\mbox{ at } R\sim 100\mbox{
TeV},
\]
$\theta$-depending correction. At this level one could expect family
non-universal couplings of the fermions with the $Z$ boson. However, this
non-universality is also generated due to an interchanging by non-zero
modes of photons or (and) $Z$ bosons. One can easily estimate the latter
in a way similar to Appendix~\ref{app:LgaugeFermi}, and finds it at the
level of
\[
g\theta_A^2\sim 10^{-2}g\gg 10^{-4}g^2.
\]
Therefore, we do not take into account the non-trivial profile of
$Z$-boson mode and treat it as a constant.

\section{Gauge--fermion Lagrangian.}
\label{app:LgaugeFermi}
Substituting Eqs.~(\ref{Eq/Pg3/5:vector}),
(\ref{Eq/Pg3/4:vector}), (\ref{Eq/Pg3/3:vector}), and
(\ref{Eq/Pg3/2:vector}) into Eq.~(\ref{Eq/Pg3/1:vector}), one finds
\[
J^{4,5}=0,
\]
\[
J^\mu =\sum \limits_{m,n=1}^{3}\rho _{mn}{\rm e}^{i\varphi(m-n)
}j_{mn}^\mu,
\]
where
\[
\rho _{mn}=\rho _{nm}=f_2(m)f_2(n)+f_3(m)f_3(n)   ,
\]
\[
j_{mn}^\mu(x) =(j_{nm})^*=a^\dag _m\bar{\sigma }^\mu a_n=\bar{\psi
}_m\gamma ^\mu\frac{1+\gamma _5}{2} \psi _n
\]
and the four-dimensional Dirac matrices are
\[
\gamma ^\mu =\left(
\begin{array}{cc}
0&\sigma ^\mu \\
\bar{\sigma }^\mu &0
\end{array}
\right), \ \ \ \gamma _5=\left(
\begin{array}{cc}
1&0\\
0&-1
\end{array}
\right) ;
\]
$\psi _n(x)=(a_n(x),b_n(x))^T$. We see that the zero fermionic modes do
not interact with the scalar fields $B_l$, $B_{lm}$, while their
interaction with the vector Kalutza-Klein tower is given by
\[
{\cal L}_6=e_6\frac{\sqrt{-G}}{R}\sum \limits_{n,m}^{3}\sum
\limits_{l=0}^{\infty }\sum \limits_{k=-l}^{l}\rho
_{mn}Y_{lk}(\theta,\varphi){\rm e}^{i\varphi (m-n)}\cdot A_\mu
^{l,k}(x)j^\mu _{mn}(x)
\]
(hereafter, we neglect higher fermionic modes whose contributions to
flavour violating processes are suppressed). In the effective
four-dimensional theory,
\begin{equation}
{\cal L}_4=\int\limits_{0}^{2\pi }\! d\varphi \int\limits_{0}^{\pi }\!
d\theta {\cal L}_6=e\cdot\sum \limits_{m,n}^{3}\sum \limits_{l=0}^{\infty
}(-1)^{\frac{(n-m)-|n-m|}{2}}E_{mn}^{l,m-n} A_\mu ^{l,(n-m)}(x)j^\mu
_{mn}(x)                ,
\label{Eq/Pg5/2:vector}
\end{equation}
where
\[
E_{mn}^{l,k}= \int\limits_{0}^{2\pi }\! d\varphi \int\limits_{0}^{\pi }\!
d\theta \sqrt{-G}\rho _{mn}Q_l^{|k|}{\rm e}^{i\varphi (m-n+k)},
\]
\begin{equation}
E_{mn}^{l,n-m}=E_{nm}^{l,n-m}= 2\pi \int\limits_{0}^{\pi }\! d\theta
\sqrt{-G}\rho _{mn}Q_l^{|n-m|},
\label{Eq/Pg5/4:vector}
\end{equation}
\begin{equation}
Q_l^{j}=(-1)^{j}\sqrt{2l+1}\sqrt{\frac{(l-j)!}{(l+j)!}} P_l^{j}(\cos\theta
).
\label{Eq/Pg5/5:vector}
\end{equation}
In Eq.~(\ref{Eq/Pg5/2:vector}), integration over $\varphi$ resulted in
selection rules which mean, in particular, that the $\varphi$-independent
zero mode of the vector field does not mediate flavour changing ($m\ne n$)
processes.

The four-dimensional charge $e$ is
\[
e=\frac{e_6}{\sqrt{4\pi }R}\cdot E^{0,0}_{nn}=\frac{e_6}{\sqrt{4\pi
}R}\cdot 2\pi \int\limits_{0}^{\pi }\! d\theta \sqrt{-g}\rho _{nn}Q_0^0 =
\frac{e_6}{\sqrt{4\pi }R}          .
\]
where we have used Eqs. (\ref{Eq/Pg3/2A:vector}), (\ref{Eq/Pg5/5:vector}).

To estimate $E_{mn}^{l,m-n}$, we note that $f_i$ are localized in the
region $\theta <\theta _A$, where $\theta _A$ is the size of the gauge
field which forms the vortex. We are working in the regime
\[
0<\theta _\Phi <\theta _A<\theta _\Psi \ll1,
\]
where $R\theta _\Phi$ is the size of the vortex on which fermions are
localised and $(R\theta _\Psi)^{-1}$ is the energy scale of the fermionic
non-zero modes (see Ref.~\cite{FLNT} for details). At $\theta \geq\theta
_A$,
\begin{equation}
f_i(n) {\sim}\frac{1}{\theta ^n}{\rm e}^{-\theta/\theta_\Psi} ,
\label{Eq/Pg6/3:vector}
\end{equation}
Therefore, the integral (\ref{Eq/Pg5/4:vector}) is saturated at
$\theta<\theta_A$, and we can use the behaviour of $P_l^m$ at the origin
($J_m$ is Bessel function of the first kind),
\[
P_l^m(\cos\theta)=(-1)^m\left[\!\left(l+\frac{1}{2}\right)\cos\frac{\theta
}{2}\right]^m\!J_m\left((2l+1)\sin\frac{\theta }{2}\right)+{\cal
O}\left(\sin^2\frac{\theta }{2}\right)\sim(-1)^ml^mJ_m(l\theta ),
\]
to find
\begin{equation}
Q_l^m\sim\sqrt{l}J_m(l\theta )\simeq\left[
\begin{array}{c}
l^{m+1/2}\theta ^m\ \ \mbox{at} \ \ l\theta _A\ll1,\\
\\
\displaystyle \frac{\cos(l\theta -1/2(m\pi) -1/4\pi )}{\sqrt{\theta }}\
\ \mbox{at} \  \ l\theta _A\gtrsim 1                 .
\end{array}
\right.
\label{Eq/Pg7/1:vector}
\end{equation}

At $l\theta _A\ll1$, one obtains, from Eqs.~(\ref{Eq/Pg7/1:vector}) and
(\ref{Eq/Pg5/4:vector}),
\[
E_{mn}^{l,m-n}\sim l^{|m-n|+1/2}\int\limits_{0}^{\pi }\! d\theta
\sqrt{-G}\rho _{mn}\theta ^{|m-n|}.
\]
This integral can be estimated in the saddle point approximation. Indeed,
$f_i(n)$ have a well localized maximum at $\theta \sim \theta _A$, so
$\sqrt{-G}\rho _{mn}$ also has a maximum near this point. Using this fact
and the normalization conditions (\ref{Eq/Pg3/2A:vector}) one finds
\[
\int\limits_{0}^{\pi }\! d\theta \sqrt{-G}\rho _{nm}\simeq 1
\]
and, thus,
\[
E_{mn}^{l,m-n}\sim l^{|m-n|+1/2}\theta _A^{|m-n|}\ \ \mbox{at} \ \ l\theta
_A\ll1.
\]

The regime $l\theta _A \gtrsim 1$ contains two cases: 1) $l\theta
_A\simeq1$, 2) $l\theta _A\gg1$. The first one can be worked out in the
same way as the $l\theta _A\ll1$ case. This is due to the fact that in
Eq.~(\ref{Eq/Pg7/1:vector}), the argument of cosine $l\theta _A\sim1$. So,
one has
\[
E_{mn}^{l,m-n}\sim \frac{1}{\sqrt{\theta _A}}\ \ \mbox{at}\ \  l \simeq
\frac{1}{\theta _A}.
\]
In the regime $l\theta _A\gg1$, however, the integral is not saturated at
$\theta \sim \theta _A$, but rather (due to quick oscillations of cosine)
at a complex value of $\theta $ --- the pole of $\ln \rho _{mn}$. Since $f_i$ have
poles at the origin at $\theta_A\to0$ (Eq. (\ref{Eq/Pg6/3:vector})), the
pole of $\ln \rho _{mn}$ should develop a non-zero imaginary part which
tends to zero as $\theta _A\to 0$. So,
\[
E_{mn}^{l,m-n}\sim {\rm e}^{-lF(\theta _A)} \ \ \mbox{at}\ \ l\theta
_A\gg1,
\]
that is the couplings to highest modes are exponentially suppressed.

\section{Rotation to physical states.}
\label{sec:Mgauge+fermi}
The mass matrix (\ref{Eq/Pg9/1:vector}) can be
diagonalized by a biunitary transformation $S_d^\dag M_DT_d=M_D^{diag}$.
One can find $S_d$ as an unitary matrix which diagonalizes $M_DM_D^\dag$:
\[
S_d^\dag M_DM_D^\dag S_d=M_D^{diag}(M_D^{diag})^\dag
\]
(the unitary matrix $T_d$ obeys $T_d^\dag M_D^\dag M_D
T_d=M_D^{diag}(M_D^{diag})^\dag$). Matrices $S_u$, $T_u$
diagonalize, in the same way, the mass matrix of up quarks $M_U$.
According to Refs.~\cite{Frere:2000dc,FLNT}, the mass matrices have the
following form,
\[M_D
\propto \left(
\begin{array}{ccc}
\delta ^4&\epsilon_d \delta ^3&0\\
0&\delta ^2&\epsilon_d \delta \\
0&0&1
\end{array}
\right),
\qquad M_U\propto \left(
\begin{array}{ccc}
\delta ^4&0&0\\
\epsilon_u \delta^3&\delta^2&0\\
0&\epsilon_u \delta&1
\end{array}\right),
\]
where a real parameter $\delta\sim0.1$ and two complex parameters
$\epsilon_u$, $\epsilon_d$ have absolute values of order $0.1$. Then, up
to the second order in $\epsilon$,
\[
S_d= \left(
\begin{array}{ccc}
1-\frac{|\epsilon_d| ^2}{2}\alpha_d^2&\epsilon_d \alpha_d&
\epsilon_d^2\beta'_d\\
-\epsilon^*_d \alpha_d&1-\frac{|\epsilon_d|^2}{2}(\alpha_d^2+\gamma_d^2)&
\epsilon_d \gamma_d\\
(\epsilon^*_d)^2\beta_d &-\epsilon^*_d \gamma_d&
1-\frac{|\epsilon_d|^2}{2}\gamma_d^2
\end{array}
\right),
\]
\[
S_u= \left(
\begin{array}{ccc}
1-\frac{|\epsilon_u| ^2}{2}\alpha_u^2&\epsilon^*_u \alpha_u&
(\epsilon^*_u)^2\beta'_u\\
-\epsilon_u \alpha_u&1-\frac{|\epsilon_u|^2}{2}(\alpha_u^2+\gamma_u^2)&
\epsilon_u^* \gamma_u\\
\epsilon_u^2\beta_u &-\epsilon_u \gamma_u&
1-\frac{|\epsilon_u|^2}{2}\gamma_u^2
\end{array}
\right),
\]
where
\[
\alpha_d=\frac{m_{22}m_{12}}{m_{22}^2-m_{11}^2}\sim\delta, \ \ \gamma_d=
\frac{m_{33}m_{23}}{m_{33}^2-m_{22}^2}\sim\delta,
\]
\[
\alpha_u=\frac{m_{11}m_{12}}{m_{22}^2-m_{11}^2}\sim\delta^3, \ \ \gamma_u=
\frac{m_{22}m_{23}}{m_{33}^2-m_{22}^2}\sim\delta^3;
\]
\[
\beta_d\sim\delta^2,\;\beta_d'\sim\delta^6,\;\beta_u\sim\delta^6,\;
\beta_u'\sim\delta^{10}.
\]
To derive the dependence of the matrices $T$ on $\epsilon$ and $\delta$ one
can use the following trick: the matrix $M_u$ transfers to $M^\dag _d$
under replacement $\epsilon_u\to\epsilon_d^*$. This means that $T_d$ have
the same structure as $S_u$ with replacement  $\epsilon_u\to\epsilon_d^*$.
From the expressions for $S$ and $T$, one can see that
$M_D^{diag}$ and $M_U^{diag}$ are real.

Let us consider now $U^{CKM}=S_u^\dag S_d$. Unphysical phases can be
removed by a transformation $F_u^\dag U^{CKM}F_d$, where
\[F_d
= \left(
\begin{array}{ccc}
{\rm e}^{i\phi^d_1}&0&0\\
0&{\rm e}^{i\phi^d_2}&0\\
0&0&{\rm e}^{i\phi^d_3}
\end{array}
\right)
\qquad F_u= \left(
\begin{array}{ccc}
{\rm e}^{i\phi^u_1}&0&0\\
0&{\rm e}^{i\phi^u_2}&0\\
0&0&{\rm e}^{i\phi^u_3}
\end{array}\right)
\]
However, $S_u$ is ``almost'' diagonal.
So, in the first approximation $U^{CKM}=S_d$
and all CKM phases can be removed by
$F_u$, $F_d$ with
$\phi^u_i=\phi^d_i\equiv\phi_i$ for $i=1,2,3$ and
$\phi_1-\phi_2=\phi_2-\phi_3={\rm arg}\epsilon_d\equiv\phi_d$.
If we take into account $S_u$, $(\phi_1-\phi_2)$ recives a
small correction ($\phi_u\equiv {\rm arg}\epsilon_u$),
\begin{equation}
\phi_1-\phi_2=\phi_d+
(\alpha_u/\alpha_d)|\epsilon_u/\epsilon_d|\sin(\phi_d+\phi_u).
\label{delta_phi}
\end{equation}
After these phase rotations and transformations to the mass basis
for fermions, the matrices of gauge interactions become
$F_d^\dag S_d^\dag A_\mu S_dF_d$ and  $F_d^\dag T_d^\dag A_\mu T_dF_d$.
Off-diagonal elements of $T_d$ are smaller than the same elements
of $S_d$, so the dominant flavour violating interactions are determined by
the matrix $\tilde{\bf A}^\mu=S_d^\dag{\bf A^\mu}S_d$:
\begin{equation}
\left(\! \! \!
\begin{array}{ccc}
{\bf A}_{11}-2{\rm Re}(\epsilon^* \alpha {\bf A}_{12})&{\bf
A}_{12}\! +\! \epsilon \alpha ({\bf A}_{11}\! \!  -\! \! {\bf A}_{22}\!
)\! -\! \gamma \epsilon^* {\bf A}_{13} &{\bf A}_{13}+\epsilon (\gamma {\bf
A}_{12}-\alpha {\bf A}_{23})\\
\\
\! {\bf A}_{12}^*\! \! +\! \epsilon^* \alpha ({\bf A}_{11}\! \! -\!
\! {\bf A}_{22})\! -\! \epsilon \gamma {\bf A}_{13}^*) & {\bf A}_{22}\!
+\! \displaystyle 2 {\rm Re}(\epsilon^*\! (\alpha {\bf A}_{12}\! -\!
\gamma {\bf A}_{23})) &{\bf A}_{23}\!\!  +\! \epsilon^* \alpha
{\bf A}_{13}\! \! +\! \! \epsilon \gamma ({\bf A}_{22}\! \! -\! \! {\bf
A}_{33}))\\
\\
{\bf A}_{13}^*+\epsilon^*(\gamma {\bf A}_{12}^*-\alpha {\bf A}_{23}^*) &
{\bf A}_{23}^*\! +\! \epsilon \alpha {\bf A}_{13}^*\!\!  +\epsilon ^*\! \!
\gamma ({\bf A}_{22} \! \! -\! \! {\bf A}_{33})& {\bf
A}_{33}+2{\rm Re}(\epsilon^* \gamma {\bf A}_{23})
\end{array}
\! \! \! \right)^{\displaystyle\mu}
\label{Eq/Pg11/1:vector}
\end{equation}
(we denoted $\alpha_d$ ($\epsilon_d$) as $\alpha$ ($\epsilon$)
for convenience and presented $S_d$ in the
first approximation on $\epsilon_d$ ).
After transformation $F^\dag_d\tilde A_\mu F_d$, the element $\tilde
A_{12}$ recives an additional phase ${\rm e}^{i(\phi_2-\phi_1)}$, which is
approximately opposite to phase of $\epsilon_d$, Eq.~(\ref{delta_phi}).
It means that the
phase of interaction through $(A_{11}-A_{12})$, which is responsible for
CP-violation in kaons (see Sec.\ref{dg=2}) is suppressed in our model:
\[
{\rm arg}(\epsilon_d {\rm e}^{i(\phi_2-\phi_1)})=
-(\alpha_u/\alpha_d)|\epsilon_u/\epsilon_d|\sin(\phi_d+\phi_u)
\]

The consideration of  $U^{CKM}$ in second order in $\epsilon$ gives
$U^{CKM}_{13}\sim\epsilon^*_u\epsilon^*_d\delta ^4$,
$U^{CKM}_{31}\sim(\epsilon^*_d)^2\delta ^2 $. So, the CKM matrix can not be
made real by any rotations by $F_u$, $F_d$, so the CP-violation in our
model arises in the same way as in the Standard Model.

\section{Effective four-fermion interactions.}
\label{sec:4fermi}
One may calculate the effective charge,
\[
g_{mn}=e^24\pi ^2\int\limits_{0}^{\pi }\!\int\limits_{0}^{\pi }\!d\theta
d\theta ' \sqrt{G(\theta )G(\theta ')}\rho _{mn}(\theta )\rho _{mn}(\theta
')\sum \limits_{l=1}^{\infty }\frac{R^2}{l(l+1)}Q_l^{|n-m|}(\theta
)Q_l^{|n-m|}(\theta' ),
\]
directly, by making use of the fact that, at $m>0$,
\[
\sum \limits_{l=m}^{\infty }\frac{1}{l(l+1)}Q_{l}^m(\theta )Q_{l}^m(\theta
')=-2G_m(\theta ,\theta '),
\]
where
\[
\left( \frac{1}{\sin\theta } \partial _\theta \sin\theta \partial _\theta
-\frac{m^2}{\sin^2\theta }\right)G_m(\theta ,\theta ')=\delta (\theta
-\theta ').
\]
Explicitly,
\[
G_m(\theta ,\theta ')=-\frac{1}{2m}\left(\tan^m\frac{\theta
}{2}\cot^m\frac{\theta '}{2}\Theta (\theta '-\theta ) +
\tan^m\frac{\theta' }{2}\cot^m\frac{\theta }{2}\Theta (\theta -\theta'
)\right).
\]
At $m=0$,
\[
\sum \limits_{l=1}^{\infty }\frac{1}{l(l+1)}Q_l^0(\theta )Q_l^0(\theta
')=-2G_0(\theta ,\theta '),
\]
where
\[
\frac{1}{\sin\theta }\partial _\theta \sin \theta \partial _\theta
G(\theta ,\theta ')=\delta (\theta -\theta ')-\frac{1}{2}
\]
and
\[
G(\theta ,\theta ')=\frac{1}{2}+\Theta (\theta '-\theta
)\ln\left(\sin\frac{\theta '}{2}\cos\frac{\theta }{2} \right)+\Theta
(\theta -\theta ')\ln\left(\sin\frac{\theta }{2}\cos\frac{\theta '}{2}
\right).
\]
Then for $m\neq n$,
\[
|g_{mn}|<8\pi ^2e^2R^2\left(\int\limits_{0}^{\pi }\!d\theta
\sqrt{-G}\rho _{mn}\tan^{|m-n|}(\theta )\right)\times
\left(\int\limits_{0}^{\pi }\!d\theta \sqrt{-G}\rho
_{mn}\cot^{|m-n|}(\theta )\right)\sim
\]
\[
e^2R^2 \theta _A^{|m-n|}\times\theta _A^{-|m-n|}=e^2R^2,
\]
which coincides with the naive estimate (\ref{Eq/Pg8/4:vector}). For
$m=n$, the result is the same (the dominant contribution comes from the
constant term in $G_0(\theta ,\theta ')$).

\end{document}